# Perfectly vertical silicon metamaterial grating couplers with large segmentation periods up to 650 nm

Jianhao Zhang,[1] Daniele Melati,[2] Yuri Grinberg,[3] Martin Vachon,[1] Shurui Wang,[1] Muhammad Al-Digeil,[3] Siegfried Janz,[1] Jens H. Schmid,[1] Pavel Cheben,[1] and Dan-Xia Xu[1,a)]

[1]*Quantum and Nanotechnology Research Centre, National Research Council, Ottawa K1A0R6, Canada.*
[2]*Centre de Nanosciences et de Nanotechnologies, CNRS, Université Paris-Saclay, Palaiseau 91120, France*
[3]*Digital Technologies Research Centre, National Research Council, Ottawa K1A0R6, Canada.*
[a)]*Author to whom correspondence should be addressed: Dan-Xia.Xu@nrc-cnrc.gc.ca*

**Abstract**: Perfectly vertical grating couplers leveraging metamaterials can achieve both high coupling efficiency and minimal back reflection using straightforward fabrication processes. The fabrication fidelity of these designs, with segmentations in both the longitudinal and transverse dimensions, hinges on the minimum feature size offered by cutting-edge fabrication technologies. In this work we present both numerical and experimental evidence that high performance devices can be obtained by using unusually large transverse segmentation periods of up to 650 nm, thereby increasing the critical feature sizes. For simple one-step full etch couplers produced on the 220 nm silicon-on-insulator platform, we demonstrate coupling efficiencies of nearly 50% in the C-band and remarkably low back reflections of -22 dB at zero-degree incidence angle. The duty cycles used in our optimized designs deviate significantly from those predicted by traditional effective medium models. Intriguingly, these discrepancies are pronounced even for modest segmentation periods such as 300 nm. Our discoveries hold promise for expanding the range of optical properties achievable in metamaterials and offer fresh insights into the fine-tuning of nanophotonic devices.

## I. INTRODUCTION

Silicon photonics plays a key role in diverse areas spanning from optical communications to quantum photonics and sensing. These domains all require interfacing between on-chip optical components and the external world. Coupling to fibers is essential for packaging and circuit-level characterization. Surface grating couplers[1-10] are commonly used due to their high flexibility and accessibility. Perfectly vertical grating couplers[5-9] are particularly sought after as they enable straightforward coupling of light to lasers, detector arrays, and multi-core fibers. However, the condition for second order diffraction is satisfied at the same time, generally producing strong back reflection to the input waveguide. Various approaches have been proposed to address this issue, with varying degrees of structural complexity and performance. Examples include employing an amorphous silicon overlay[9,10] or a slanted grating structure[11]. A CMOS-compatible approach uses unit cells consisting of a pillar and a L-shaped segment, resulting in high vertical coupling efficiency (CE) and low back reflection (BR)[5-8]. However, the complexity of these unit cells reduces the minimum feature size (MFS) compared to the commonly used grating couplers that use an approximately 10-degree emission angle for suppressing back reflection, as well as requiring two etch steps. These factors pose challenges for fabrication[12-14], leading to poor fidelity and reproducibility.

In our previous work, we tackled the MFS issue in vertical grating couplers by incorporating metamaterials with segmentation periods of up to 450 nm into the unit cells, and enabled MFS > 100 nm in a two-step etched grating coupler, with a small penalty in the coupling efficiency[7]. Similar to a substantial volume of existing research, these couplers were designed for SMF28 fibers with a 10.4 μm mode size. However, it is highly desirable to further enlarge or at least preserve the feature sizes while maintaining high performance in couplers for a range of fiber mode sizes that require different scattering strength.

In the component design process, metamaterials[15] formed by periodically segmenting the core (silicon) medium are conventionally represented by an effective material index $n_M$, controlled only by the segmentation period ($\Lambda_T$) and the duty cycle (DC), according to the effective medium theory (EMT). Detailed analysis has been conducted for the 'laminar' case of infinitely extended stratified layers, or the 'slab' case where the finite silicon device layer thickness is taken into account[16,17]. The outcome of the two models differs only slightly, particularly when $\Lambda_T$ is small relative to the wavelength. The $\Lambda_T$ should stay below a certain level to avoid unwanted diffraction. A wide range of $n_M$ is desirable in many cases for enlarging the feature sizes in the longitudinal direction. It is also very useful for designing weak gratings to achieve larger mode size. However, the achievable MFS in fabrication ultimately constrains the available range of effective indices and device yields.

In this work, we present a significant advance in the understanding of metamaterial design that enables the use of markedly expanded periods and DC ranges, and apply the methodology to improve perfectly vertical grating coupler designs. In particular, we show that conventional effective medium models do not correctly predict the duty cycle of metamaterials that are embedded in more complex 3D nanostructures. These discrepancies can completely invalidate simulation results for larger periods using $n_M$ to represent metamaterials. Remarkably, even at rather small $\Lambda_T$ of 200 nm, the EMT models fail for certain device designs, such as Design B described below. Through rigorous 3D FDTD simulations and experimental investigations, we provide compelling evidence that $\Lambda_T$ as large as 650 nm still yields high-performance grating couplers for the design configurations described below, but with modified duty cycles.

## II. DESIGN

### A. Initial design procedure and the inadequacy of the effective medium predictions

We have investigated several grating configurations for coupling to a wide range of fiber modal diameters, that require different combinations of $n_M$ and segment sizes. The general findings are similar. In this report, we focus on single-step fully etched structures in a 220 nm SOI platform, designed for standard SMF28 fibers in the C-band with a mode size of 10.4 µm, as it enables straightforward experimental characterization. The buried oxide and the upper cladding oxide layers are 2 µm and 2.2 µm thick, respectively. Their refractive indices are considered to be the same in simulation. The input slab waveguide of 15 µm in width receives light in the fundamental mode with transverse electric (TE) polarization. This light is diffracted by the grating coupler structure and captured by the vertically positioned SMF28 fiber.

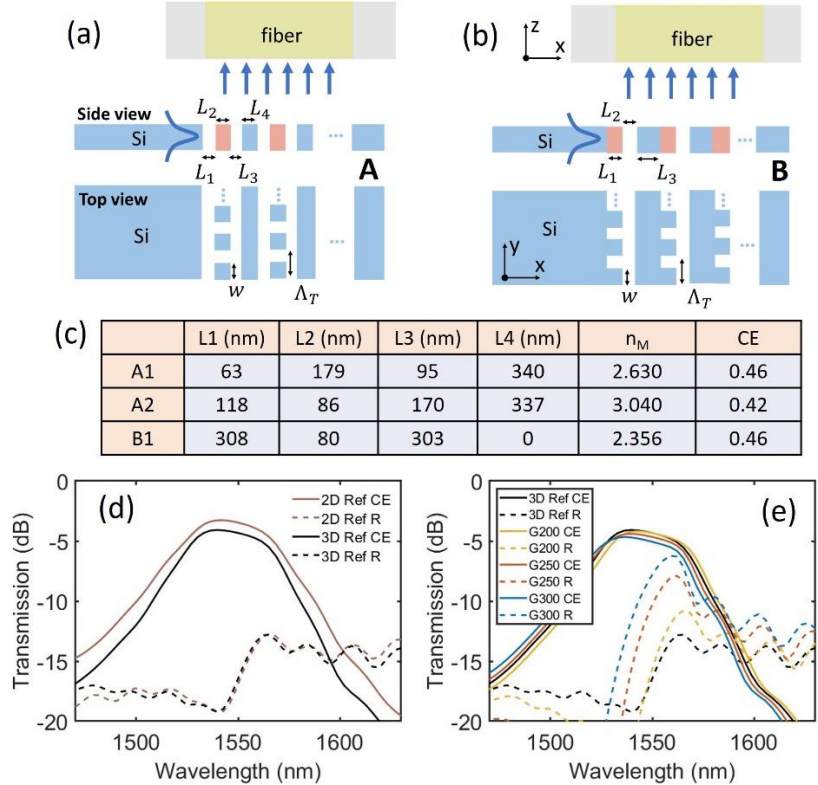

FIG. 1. (a) and (b) Schematic of the type A and B grating structures. The metamaterial section with segmentation is highlighted in orange. (c) Device structural parameters and predicted peak coupling efficiency (CE) using 2D simulations. (d) CE and reflection R of design A1 calculated using 2D and 3D simulations as a function of wavelength, with the metamaterial section represented as an effective medium of index $n_M$ in both cases. (e) CE and R of design A1 using 3D simulation with segmentation as compared to the effective medium case. The legends G200, G250 and G300 indicate the segmentation periods $\Lambda_T$ used.

Two kinds of unit-cell configurations are presented [Fig. 1(a) and Fig. 1(b)] to illustrate that the behavior of the metamaterial is influenced by its surroundings. Type A consists of 4 longitudinal sections, with a metamaterial section and a silicon section separated by oxide slots. Type B comprises 3 sections, including a metamaterial section, an oxide slot, and a silicon section. They are single-step etch variants of the previously studied designs[7,8]. Compared to the L-shaped structure used in Ref. [4, 5] grating blazing is replaced by a simple silicon section of uniform thickness. The directionality is therefore reduced, but with the benefit of simplified fabrication. The initial design procedure follows the global mapping methodology developed previously[8]. Assuming that the metamaterial can be represented by an effective medium with a refractive index $n_M$, we perform 2D finite difference time domain (FDTD) simulations to compute the coupling efficiency (CE) and back reflection (R) at the optimized fiber height. The longitudinal (i.e., along the x-direction) section lengths ($L_i$) and $n_M$ are randomly sampled and adjusted using gradient descent to optimize CE at 1550 nm wavelength. The sampling range was constrained with MFS≥ 60 nm. The principal component analysis (PCA) is further employed to identify the hyperplane capturing the subspace of high-performance designs (see supplementary material). The resulting maps allow the selection of designs with a balance between CE and R for further investigation. Two type A designs (A1 and A2) of varied dimensions and index, and one type B design are selected [Fig. 1(c)]. The longitudinal (diffracting) period is in the range of 670 nm to 715 nm. Since the optimization was carried out at a single wavelength, the resulting design may show a peak wavelength deviated from 1550 nm. This shortcoming can be resolved numerically in future work.

We first investigate the validity of the 2D FDTD simulation results for design selection, by comparing the 2D and 3D simulations representing the metamaterial in both cases with a material index $n_M$ (referred as 2D and 3D Ref), without any segmentation in the transverse direction. In the 3D case, this equivalent material is implemented as a strip of $L_i \times 15$ µm in size in the x-y plane. As shown in Fig. 1 (d) for design A1, the back-reflections are nearly identical between 2D and 3D simulations, with R around -20 dB at the CE peak wavelength and the worst around -12 dB at 1565 nm. The slight decrease in CE for the 3D simulation, from 0.47 to 0.39, is due to the finite modal width in the transverse (y) direction. These results indicate that design selection based on 2D simulations would be valid if the effective medium model gave accurate predictions. The results obtained from the 3D effective medium simulations (3D Ref) are used as a reference baseline for further analysis.

Subsequently, 3D simulations were conducted using the full 3D segmented structures. In this initial step, the duty cycle was obtained using the slab model calculation for a given period and the desired metamaterial index[16]. This has been the typical design procedure for devices using metamaterials. In Design A, the wavelength dependence of CE and R for periods of 200, 250, and 300 nm is presented in Fig. 1(e). For the case of $\Lambda_T = 200$ nm (marked as G200), the CE closely aligns with the predictions of the slab model, suggesting that segmented structures with such small periods can be adequately represented by the effective medium approximation. However, as the period increases, the reflections (dashed lines) notably increase, even at a small period of 300 nm. For larger periods, this trend persists and the back-reflection rapidly surpasses the power in upward diffraction (see supplementary material). Clearly, the designs require further refinement.

B. Device optimization using 3D FDTD to overcome the challenges in the index prediction

We then conducted a 3D optimization process to explore a wide range of transverse segmentation periods to avoid unattainable feature sizes while minimizing back-reflections. The process is implemented as follow: starting from the DC predicted by the slab model for each $\Lambda_T$ for the required index, a scan on the DC was performed (scan step≤ 0.025) by balancing the figures-of-merit until the CE, R and most importantly the peak wavelength and the spectrum shape to achieve a close match to the reference, namely the design using an index-defined homogeneous medium [3D Ref in Fig. 1(d)]. We take this matching as an indication that the desired $n_M$ is achieved. It's import to note, however, that this matching process is approximate and should be regarded only as an indicator rather than a precise solution.

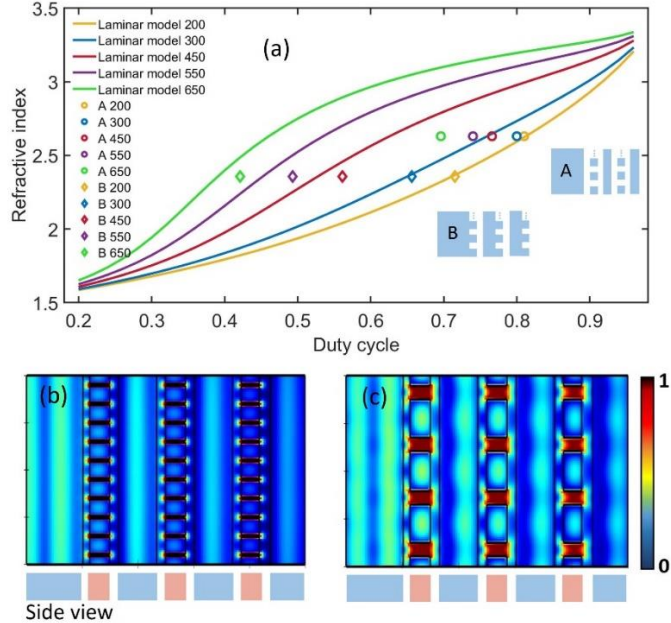

FIG. 2. (a) Effective refractive index of the metamaterial as a function of the duty cycle for different transverse periods, as predicted using the laminar model (curves) and optimized in 3D-FDTD simulation (symbols). Circles are for design A1 and diamonds are for design B. (b) and (c) Optical near-field distribution of design A1, with period of 200 nm and 550 nm, respectively. Geometric boundaries are outlined by black solid lines.

The conclusive comparison on the refractive indexes $n_m$ and the duty cycle DC is shown in Fig. 2(a). Solid lines show $n_M$ versus DC predicted using the laminar model. The 3D optimized duty cycle (DC$_{3D}$) of design A1 and B segmented structures are represented as circles and diamonds, respectively. The segmentation periods are distinguished by the colors. One can observe that for design A1 at $\Lambda_T$= 200 nm, the laminal model and 3D segmentation simulation indicate nearly the same duty cycle of 0.81 for $n_M$= 2.63. However, as $\Lambda_T$ increases to 300 nm, the 3D full simulations start to deviate. We obtained DC$_{3D}$= 0.800, 0.766, 0.740 and 0.696 for configurations with $\Lambda_T$ of 300 nm, 450 nm, 550 nm and 650 nm, respectively.

To gain physical insight, we compared the optical electric near field distributions of design A1 over two periods. The effective medium theory (EMT) assumes an average-field representing light propagation in segmented metamaterials, with an effective index $n_M$[17]. However, the actual metamaterial structures exhibit highly nonuniform fields [Fig. 2(b) and 2(c)]. In adjacent homogeneous Si and SiO$_2$ sections, the field remains uniform in the transverse y-direction for small periods [$\Lambda_T$= 200 nm in Fig. 2(b)], behaving like an 'average field' transmitted from the metamaterial region. However, as $\Lambda_T$ increases [Fig. 2(c)], the fields in the Si and SiO$_2$ sections are strongly modulated also in the y- direction, as imposed by the electro-magnetic boundary conditions. While EMT correctly predict light behavior in extended metamaterial, it falls short in composite structures like our gratings[3], especially as $\Lambda_T$ grows. This also explains the marked differences found between type A and B configurations, as summarized in Fig. 2(a), since they experience very different boundary conditions.

## III. EXPERIMENTAL DEMONSTRION OF THE OPTIMIZED DESIGNS

The predicted performance of design A1 with 3D optimized duty cycles for different periods is shown in Fig. 3(a). One can see that all configurations give similar performance on coupling efficiency. Devices were patterned using electron beam lithography, then followed by inductively coupled plasma etching for fabrication. Scanning electron microscope (SEM) image of a fabricated device with 650 nm period is shown in Fig. 3(b), while the optical performance of fabricated devices is shown in Fig. 3(c) and 3(d) for comparison. Two identical grating couplers are connected with a straight waveguide 450 nm wide and 2.5 mm long. The light input and output are coupled to two polarization-maintaining cleaved fibers with anti-reflection coating, placed vertically above the chip. The reported results are normalized for a single grating coupler, with the waveguide loss (1.5 dB/cm) subtracted.

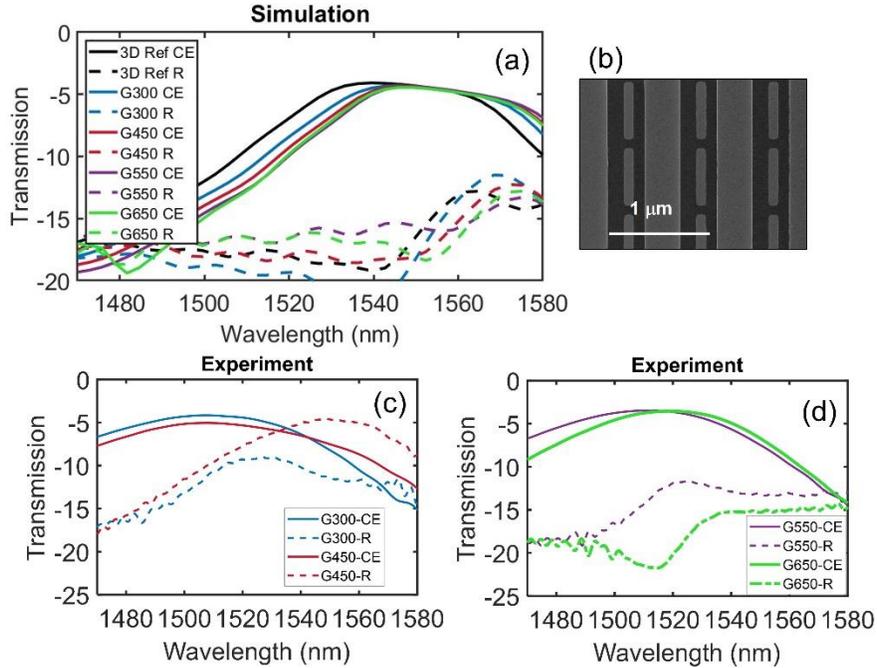

FIG. 3. (a) Coupling efficiency (CE) and reflection (R) of design A1 calculated using 3D simulation with segmentation. The duty cycle is separately optimized for each segmentation period. (b) SEM image of fabricated device with 650 nm period. (c) and (d) Experimental results of design A1 with different periods $\Lambda_T$.

The original transmission spectra display a fringe pattern as a result of the Fabry-Perot cavity formed by back reflections from the two grating couplers (see supplementary material). The results shown in Fig. 3(c) and 3(d) were extracted from the raw spectra using a Fourier transform (see supplementary material) based method[18,19]. The back reflection is strong for devices with $\Lambda_T$ of 300 nm and 450 nm, with a wavelength dependence similar to the prediction, but with a higher magnitude. These outcomes are partially attributed to the feature size and shape deviations in fabricated devices, which are more severe in small features, as observed in SEM imaging. The coupling efficiency with 550 nm and 650 nm periods is comparable to that of the 300 nm period, similar to the simulation results in Fig. 3(a). However, the back reflections are significantly lower. Particularly, the coupler with 650 nm period shows a peak CE of 0.48 (-3.2 dB) at 1515 nm, and extracted R below -22 dB. Here the features are well reproduced as shown in Fig. 3(b). The A2 design shown in Fig. 1(c) is also investigated experimentally to highlight the achievable high index of $n_M$= 3.04, which led to enlarged section dimensions in the longitudinal direction. The device achieved a -4.2 dB coupling efficiency and low back reflection of -25 dB (supplementary material). These two devices were selected from the hyperplane of good designs shown in Fig. S1. With this new found knowledge that high metamaterial index is indeed achievable when using large $\Lambda_T$, different designs can be selected from the hyperplane that may better satisfy different requirements

To explore different possible grating coupler configurations and the behavior of metamaterials in different environments, we also investigated the type B structure shown in Fig. 1. A biasing of the longitudinal dimensions of the structure was found necessary in DC scanning to yield reasonably good designs. The modified nominal dimensions are $[L_1, L_2, L_3]$ = [308, 100, 283] nm. Even with a 200 nm period, the effective medium model failed to give reliable predictions for this design, particularly for back reflection (see Fig. S3(a)). The best performance is predicted with a period of 450 nm, as shown in Fig. 4(a). The near-field electric field distribution (see supplementary material) is significantly modified in both the transverse and longitudinal directions. These results show again that optical structures consisting of different combinations of homogeneous and metamaterials cannot be modelled using simple effective medium approximations. The measured optical performances of the fabricated design B devices are shown in Fig. 4(c) and 4(d). The strong fringes in the long wavelength ends are numerical artifacts of the Fourier extraction process (see raw data in Fig. S5). The general trends are similar to the simulation except that the coupling efficiency of 450 nm (-5.3 dB) is exceeded slightly by that of 550nm (-4.2 dB), while the reflection is very low for the 650 nm case with comparable CE (-5.6 dB).

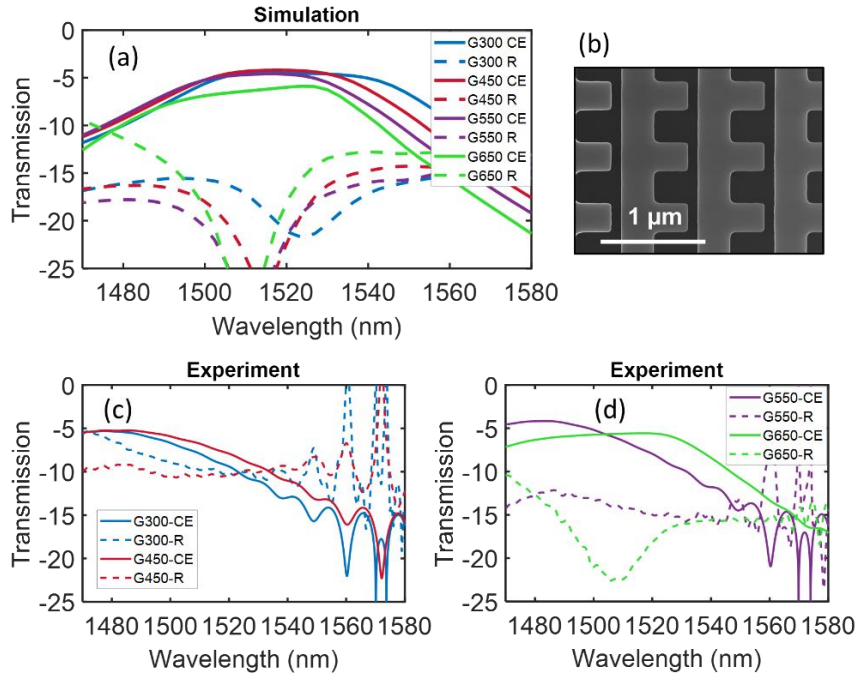

FIG. 4. (a) Coupling efficiency (CE) and reflection (R) of the design B calculated using 3D simulation with segmentation. The duty cycle is separately optimized for each segmentation period. (b) Scanning electron microscope image of fabricated device with 650 nm period. (c) and (d) Experimental results of type-B devices of different periods.

## IV. DISCUSSIONS AND CONCLUSIONS

In summary, we have successfully integrated metamaterials with segmentation periods extending to an uncommon range of 650 nm, resulting in perfectly vertical grating couplers that offer high performance and larger feature sizes. We further show that the effective medium models proved inadequate for predicting outcomes in complex structures for even modest segmentation periods such as 300 nm. Using 3D FDTD simulations we effectively identified optimized duty cycles, which are not solely dictated by the segmentation period but are also significantly influenced by external conditions such as segment length and surrounding structure. Our experiments achieved impressive coupling efficiencies of nearing 50% for single-step etch couplers, with reduced back reflections at -22 dB, and transverse feature sizes surpassing 150 nm. These coupling efficiencies already surpassed the conventional single step-etched grating coupler with apodization (35% efficiency at 10° angle)[2], while achieving perfectly vertical emission and low back-reflection, in a simple, full-etch configuration. These results provide opportunities for further device optimization, leveraging a wider metamaterial parameter range for both apodization and larger mode size gratings. These advancements signify a leap forward for silicon photonic grating coupler design, regarding both performance and manufacturability.

Since our current duty cycle scanning method is heavy in computation load, we proposed using a neural network surrogate model to capture the intricate relationships between metamaterial optical properties and their environment[20]. Implementing an accurate 2D model would make device optimization methods, as outlined previously[8], practical for intricate structures as well. This strategy holds promise for other device configurations, further amplifying our capacity to engineer advanced nanophotonic devices that are robust against fabrication uncertainties.

## ACKNOWLEDGMENTS

Dr. Daniele Melati acknowledges the support from European Research Council (ERC) project BEAMS (101041131).

# Supplementary Information

## A: Design hyperplane projected using PCA method

Once several promising designs are obtained using 2D FDTD simulations, principal component analysis (PCA) is employed to identify the hyperplane capturing the subspace of high-performance designs. This hyperplane is then mapped for both CE and BR. As an example, the PCA maps on CE and BR of design B is shown in Fig. S1(a) and (b), respectively. Note that the CE displayed in the maps is calculated as the field overlap with a Gaussian beam of MFD of 10.4 um at 2 um above the silicon layer. These values are smaller than the CE with a SMF28 fiber at an optimized height.

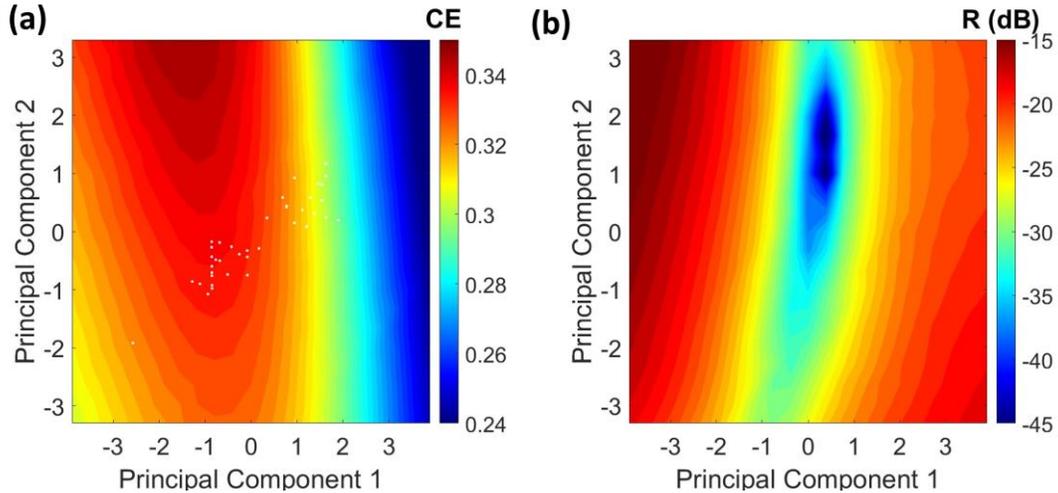

Fig. S1. (a) 2D FDTD prediction of coupling efficiency to an SMF-28 fiber for A-type design, as a function of the first two principal components. The white points indicate the local optima in the original search. (b) Prediction of back reflection.

## B: 3D Simulation settings

A mode source is launched in the feed waveguide, and a single-mode fiber with a mode size of 10.4 µm is placed on top of the grating coupler to collect the diffracted light output. The position of the fiber is optimized to achieve maximum coupling efficiency, and is centered at 3.5 µm from the beginning of the grating with the fiber face at a height of 1.1 µm above the upper cladding for all discussed designs. A mesh size of 20 nm is used in both 2D and 3D simulations in all directions. The simulation volume is of 27 µm × 27 µm × 9 µm large, assuring the inclusion of the full fiber core and substantial amount of the fiber cladding to obtain a good fiber mode without any profile truncation. Perfectly matched layers are used in all 6 boundary surfaces to avoid reflections.

## C: Additional simulation results of design A1

Additional simulation results for design D1 with the duty cycles predicted using the effective medium slab model, using a segment with an index of 2.630, are reported in Fig. S2(a, b). As reported in the main text, the grating performance deviates from that using a strip of homogeneous medium to represent the metamaterial section ($L_2$), even at a small period of 300 nm. As the period increases, the reflections (dashed lines) notably increase. For larger periods, this trend exacerbates and the back-reflection rapidly surpasses the power in upward diffraction. These results are compared with the grating performance obtained by optimizing the duty cycle using 3D FDTD shown in Fig. S3(c), where high performance is achieved for periods as large as 650 nm.

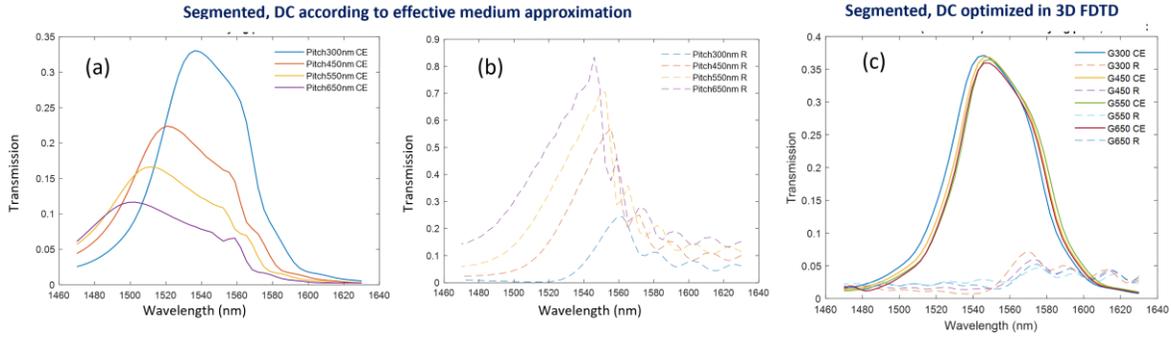

Fig. S2. 3D FDTD prediction of grating coupler performance for design A1, for segmentation periods of 300 nm, 450 nm, 550 nm and 650 nm. (a) Coupling efficiency (CE) to an SMF-28 fiber as a function of the wavelength, with the duty cycles predicted using the effective medium slab model. (b) The corresponding back reflection (R) as a function of the wavelength. (c) The CE and R predicted for designs with the segmentation duty cycle optimized for each period using 3D FDTD simulations.

## D: Additional simulation results of design B

The 3D simulation results using the effective medium (using a segment with an index of 2.356) are compared with the segmented structure with a period of 200 nm in Fig. S3(a). In this case, even at this small period well below the expected diffraction limit, the segmented structure does not behave as an effective medium. Scanning the duty cycle using the original section dimensions as listed in Table 1 also failed to produce good designs. The optical near-field distributions for periods of 200 nm and 550 nm are shown in Fig. S3(d, e), similar to that reported in Fig. 2(b, c) for design A1. As can be observed, the electric field extends significantly in the longitudinal direction even for the 200 nm period structure. For the 550 nm period structure, the E-field in the silicon segment is modulated also in the transverse direction, no longer exhibiting the local uniformity observed in the 200 nm period structure. It is then not surprising that the optimized grating designs deviate from the dimensions based on the effective medium approximation in both the transverse and longitudinal directions.

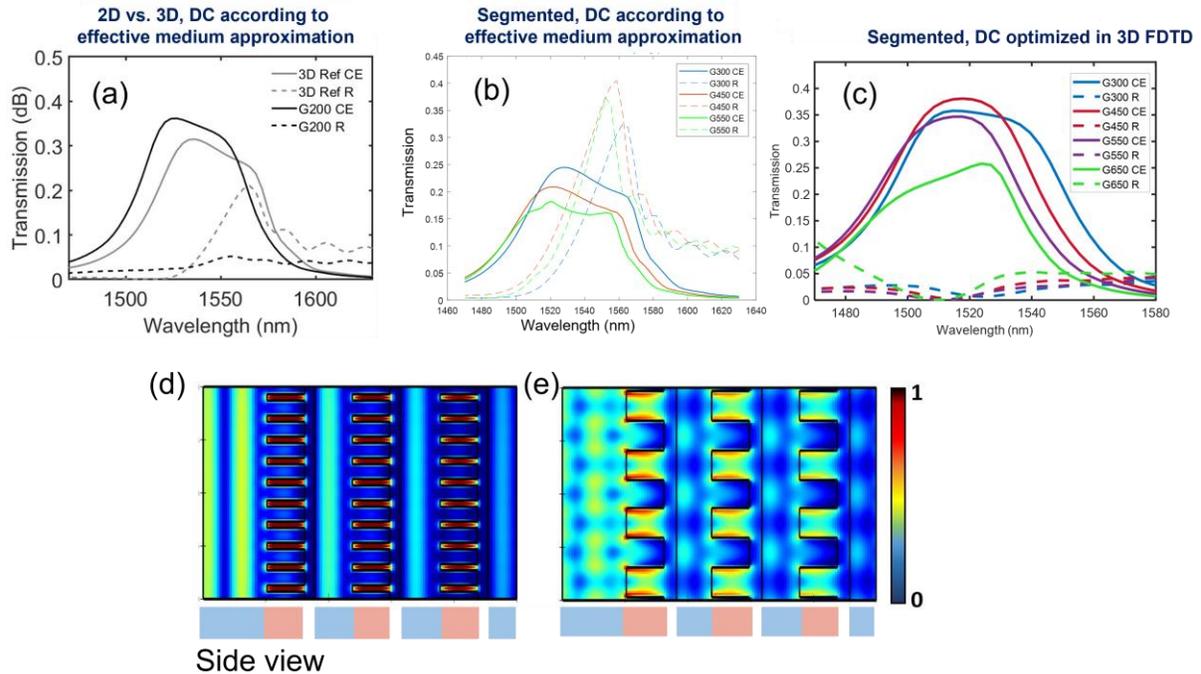

Fig. S3. (a) 3D FDTD prediction of grating coupler performance for design B. Coupling efficiencies (CE, solid lines) and reflections (R, dash lines) of design are predicted using homogeneous medium and segmentation at 200 nm period, repsectively, for comparison. (b) Coupling efficiency and back reflection as a function of the wavelength for larger segmentation periods, with duty cycles predicted using the effective medium slab model. (c) The CE and R predicted for designs with the segmentation duty cycle optimized for each period from 300 nm to 650 nm, using 3D FDTD simulations. (d, e) Optical near-field distribution of B design, with period of 200 nm and 550 nm, respectively. Geometric boundaries are outlined by black solid lines.

**E: Experimental measurement raw data of design A1 and B**

The following figures report the experimental results obtained for devices listed in the table in Fig. 1 (design A1, A2 and design B). The reported results are normalized for a single grating coupler, with the waveguide loss (1.5 dB/cm) subtracted. The processed data for design A1 and B are reported in Fig. 3 and Fig. 4 of the main text. For design A2, the effective medium index of 3.04 is rather high. Devices for smaller segmentation periods showed poor CE and very high R. Nonetheless, a large period of 650 nm makes such high index attainable, providing a good coupling efficiency of -4.2 dB.

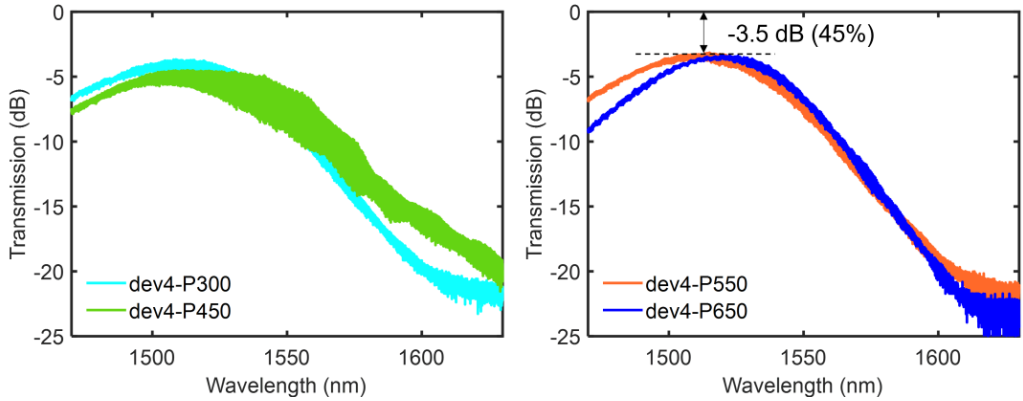

Fig. S4. Experimental performance of design A1 devices, with the duty cycles optimized in 3D FDTD. The corresponding extracted data of the coupling efficiency and back reflection using a Fourier transform method are shown in Fig. 3 of the main text.

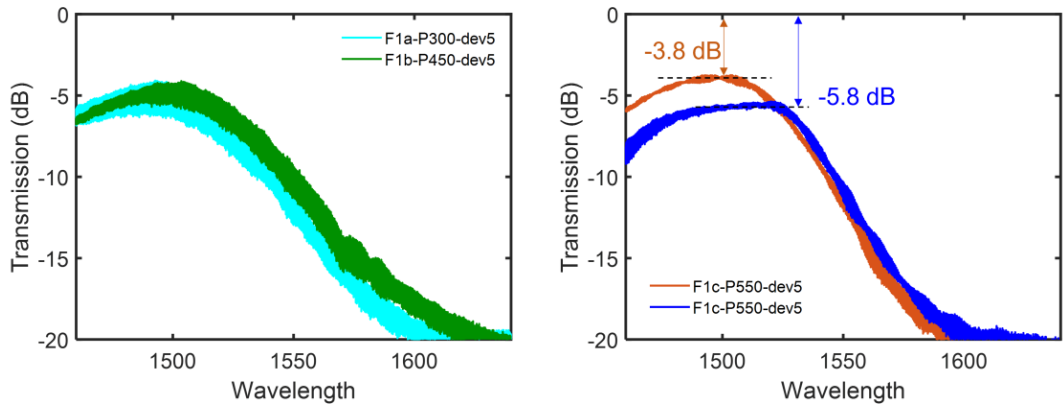

Fig. S5. Experimental performance of design B devices, with the duty cycles optimized in 3D FDTD. The extracted data of the coupling efficiency and back reflection using a Fourier transform method are shown in Fig. 4 of the main text.

**F: Method to extract the back reflection**

The back-reflections generated by the gratings have been determined from the raw experimental data using the Optical Frequency Domain Reflectometry technique [1-3]. By Fourier transforming the measured transmission spectra, it is possible to identify resonant cavities that are present in the device under test, such as the weak Fabry-Perot cavity generated by the back reflections of the input and output gratings. These cavities appear in the transformed domain as harmonic peaks whose amplitude depends on the total cavity losses and the reflectivity of the gratings. Cavity losses include both grating transmission (directly measured) and waveguide propagation losses, that have been normalized through dedicated spiral-based test structures. Cavity loss normalization allows to retrieve the desired grating back reflection level.

## G: Simulation and experimental results of A2 design

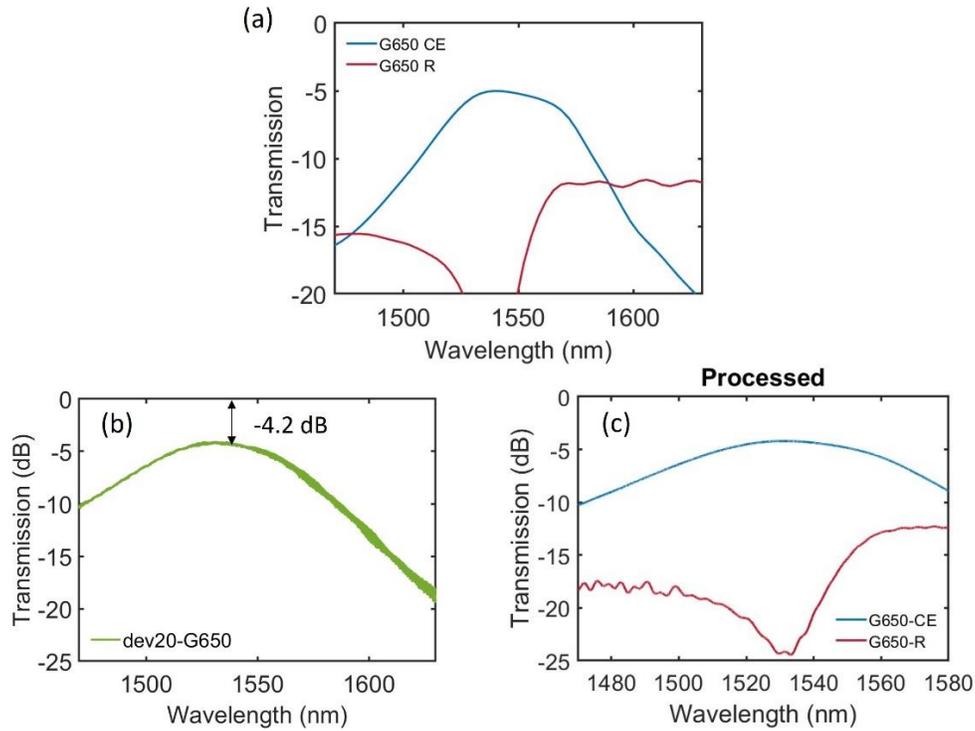

Fig. S6. (a) Coupling efficiency (CE) and reflection (R) of design A2 calculated using 3D simulation with segmentation. Experimental performance of A2 device with a 2D prediction of $n_{SWG} = 3.04$, for 650 nm transverse period. (b) Measured transmission coupled using SMF28 fibers, normalized to a single grating coupler. (c) Processed data of extracted coupling efficiency and back reflection using a Fourier Transform method.